\documentclass[aps,prl,twocolumn,showpacs,superscriptaddress]{revtex4-1}  

\usepackage[pdftex]{graphicx}
\usepackage{grffile}
\usepackage{epstopdf}

\usepackage{bm}        
\usepackage{amssymb}   

\begin{document}

\title{Carrier-envelope phase control over pathway interference in strong-field dissociation of H$_2^+$}

\author{Nora G. Kling}%
\affiliation{J. R. Macdonald Laboratory, Department of Physics, Kansas State University, Manhattan, KS 66506, USA}
\author{K. J. Betsch}%
\affiliation{J. R. Macdonald Laboratory, Department of Physics, Kansas State University, Manhattan, KS 66506, USA}
\author{M. Zohrabi}%
\affiliation{J. R. Macdonald Laboratory, Department of Physics, Kansas State University, Manhattan, KS 66506, USA}
\author{S. Zeng}%
\affiliation{J. R. Macdonald Laboratory, Department of Physics, Kansas State University, Manhattan, KS 66506, USA}
\author{F. Anis}%
\altaffiliation[Present address: ]{Department of Biomedical Engineering, Washington University in St. Louis, St. Louis MO 63130, USA}
\affiliation{J. R. Macdonald Laboratory, Department of Physics, Kansas State University, Manhattan, KS 66506, USA}
\author{U. Ablikim}%
\affiliation{J. R. Macdonald Laboratory, Department of Physics, Kansas State University, Manhattan, KS 66506, USA}
\author{Bethany Jochim}%
\affiliation{J. R. Macdonald Laboratory, Department of Physics, Kansas State University, Manhattan, KS 66506, USA}
\author{Z. Wang}%
\altaffiliation[Present address: ]{TEDA Applied Physics School, Nankai University, Tianjin 300457, China}
\affiliation{J. R. Macdonald Laboratory, Department of Physics, Kansas State University, Manhattan, KS 66506, USA}
\author{M. K\"{u}bel}%
\affiliation{Max-Planck-Institut f\"{u}r Quantenoptik, 85748 Garching, Germany}%
\author{M. F. Kling}%
\affiliation{J. R. Macdonald Laboratory, Department of Physics, Kansas State University, Manhattan, KS 66506, USA}
\affiliation{Max-Planck-Institut f\"{u}r Quantenoptik, 85748 Garching, Germany}%
\author{K. D. Carnes}%
\affiliation{J. R. Macdonald Laboratory, Department of Physics, Kansas State University, Manhattan, KS 66506, USA}
\author{B. D. Esry}%
\affiliation{J. R. Macdonald Laboratory, Department of Physics, Kansas State University, Manhattan, KS 66506, USA}
\author{I. Ben-Itzhak}%
\affiliation{J. R. Macdonald Laboratory, Department of Physics, Kansas State University, Manhattan, KS 66506, USA}

\date{\today}

\begin{abstract}
The dissociation of an H$_2^+$ molecular-ion beam by linearly polarized, carrier-envelope-phase-tagged 5\,fs pulses at 4$\times10^{14}\,$W/cm$^2$ with a central wavelength of 730\,nm was studied using a coincidence 3D momentum imaging technique. Carrier-envelope-phase-dependent asymmetries in the emission direction of H$^+$ fragments relative to the laser polarization were observed. These asymmetries are caused by interference of odd and even photon number pathways, where net-zero photon and 1-photon interference predominantly contributes at H$^+$+H kinetic energy releases of 0.2\,--\,0.45\,eV, and net-2-photon and 1-photon interference contributes at 1.65\,--\,1.9\,eV. These measurements of the benchmark H$_2^+$ molecule offer the distinct advantage that they can be quantitatively compared with \textit{ab initio} theory to confirm our understanding of strong-field coherent control via the carrier-envelope phase.
\end{abstract}

\pacs{XXX}
\maketitle

One ultimate goal of ultrafast, strong-field laser science is to coherently control chemical reactions \cite{Brumer:1992, Zewail:2000, Shapiro:2003}. A prerequisite to achieving this goal is to understand the control mechanisms and reaction pathways. To this end, tailoring the electric field waveform of few-cycle laser pulses to control reactions and uncover the underlying physics has become a powerful tool \cite{Rabitz:2000, Hoff:2011, Kling:2013}. It has been applied to the dissociative ionization of H$_2$ and its isotopologues \cite{Kling:2006, Kling:2008, Kremer:2009, Fischer:2010, Znakovskaya:2012, Xu:2013} and has recently been extended to more complex diatomic molecules, such as CO \cite{Znakovskaya:2009, Liu:2011, Betsch:2012}, and to small polyatomic molecules \cite{Xie:2012, Mathur:2013}.

Conceptually, one of the most basic features of a few-cycle laser pulse to control is the carrier-envelope phase (CEP). When the laser's electric field is written as $E(t)=E_0(t)\cos(\omega t + \phi)$, $E_0(t)$ is an envelope function, $\omega$ is the carrier angular frequency, and $\phi$ is the CEP. In fact, all of the few-cycle waveform experiments cited above used the CEP as the control parameter.

For example, Kling $\textit{et al.}$ used 5\,fs, 1.2$\times$10$^{14}\,$W/cm$^2$ pulses with stabilized CEP to dissociatively ionize D$_2$ and found asymmetries in the emission direction of D$^+$ ions for kinetic energy releases (KER) above 6\,eV \cite{Kling:2006, Kling:2008}. The diminished dissociation signal in a circularly polarized laser field indicated that recollision played a role. Recollision entails a tunnel-ionized electron undergoing a collision with its parent ion after acceleration by the oscillating laser field \cite{Corkum:1993, Kulander:1993}. The energy exchange between the laser-driven electron and the parent ion can promote the D$_2^+$ to the 2\textit{p$\sigma_u$} excited state. Coupling of the 2\textit{p$\sigma_u$} and 1\textit{s$\sigma_g$} states \cite{Bransden:2003} on the trailing edge of the laser pulse during the dissociation of D$_2^+$ was suggested as the explanation for the CEP-dependent asymmetry \cite{Kling:2006, Kling:2008}.

Another example comes from Kremer \textit{et al.} who exposed an H$_2$ target to 6\,fs, 4.4$\times$10$^{14}\,$W/cm$^2$ CEP-stabilized laser pulses and observed asymmetries for KER values between 0.4 and 3\,eV \cite{Kremer:2009} --- energies they attributed to bond softening (BS) \cite{Bucksbaum:1990} and not electron recollision, which has higher KER. They proposed that the initial ionization of H$_2$ generates a coherent wavepacket in H$_2^+$ that propagates to internuclear distances where the 1\textit{s$\sigma_g$} and 2\textit{p$\sigma_u$} states can be coupled by the tail end of the laser pulse \cite{Kremer:2009, Fischer:2010}. Bond-softening was recently found to play an even larger role in the CEP control of the dissociative ionization of D$_2$ at mid-infrared wavelengths \cite{Znakovskaya:2012}.

A wealth of theoretical studies have appeared to qualitatively interpret the main features of the CEP control in these experiments. All have modeled the ionization step and only treated the resulting H$_2^+$ explicitly using either the time-dependent Schr\"{o}dinger equation (TDSE) \cite{He:2007, He:2008, Geppert:2008, Znakovskaya:2012} or semi-classical calculations \cite{Tong:2007, Graefe:2007, Geppert:2008, Kelkensberg:2011}, assuming an initial Franck-Condon vibrational wavepacket created by the ionization of H$_2$ within the laser pulse. Moreover, due to the difficulty of treating the ionization and recollision steps, they have not yet been included in any \textit{ab initio} calculations. Therefore, quantitative agreement of accurate theoretical results and experimental data has so far been missing.

In contrast, by studying an H$_2^+$ molecular ion target, the need to model the ionization step is avoided. And, with only a single electron, recollision cannot play a role. Furthermore, state-of-the-art H$_2^+$ calculations including nuclear rotation and intensity averaging \cite{Anis:2009, Anis:2012} can now provide a nearly exact description of strong-field dissociation so long as ionization remains negligible \cite{SI2}. Thus, quantitative agreement between theory and experiment for H$_2^+$ should be attainable.

In fact, CEP control over molecular dissociation was first proposed theoretically by Roudnev \textit{et al.} for H$_2^+$ in anticipation of experiments \cite{Roudnev:2004}. Unfortunately, the low density of an ion-beam target coupled with the technical difficulties of long-time CEP stabilization have so far prevented these benchmark measurements.

Taking advantage of recent progress in phase-tagging \cite{Johnson:2011} to overcome these difficulties, we report in this Letter the measurement of CEP-dependent spatial asymmetries in the dissociation of H$_2^+$, from an ion beam, which we quantitatively compare with \textit{ab initio} theory. Roudnev and Esry have shown that these CEP effects are due to interference of different photon number pathways to 1\textit{s$\sigma_g$} and 2\textit{p$\sigma_u$} final states, having opposite nuclear parity, whose relative phase is controlled by the CEP \cite{Roudnev:2007}. Therefore, these results are a clear demonstration of strong-field coherent control.

\begin{figure} [t]
\includegraphics[width=0.48\textwidth,trim=0 0 0 0,angle=0,clip=true]{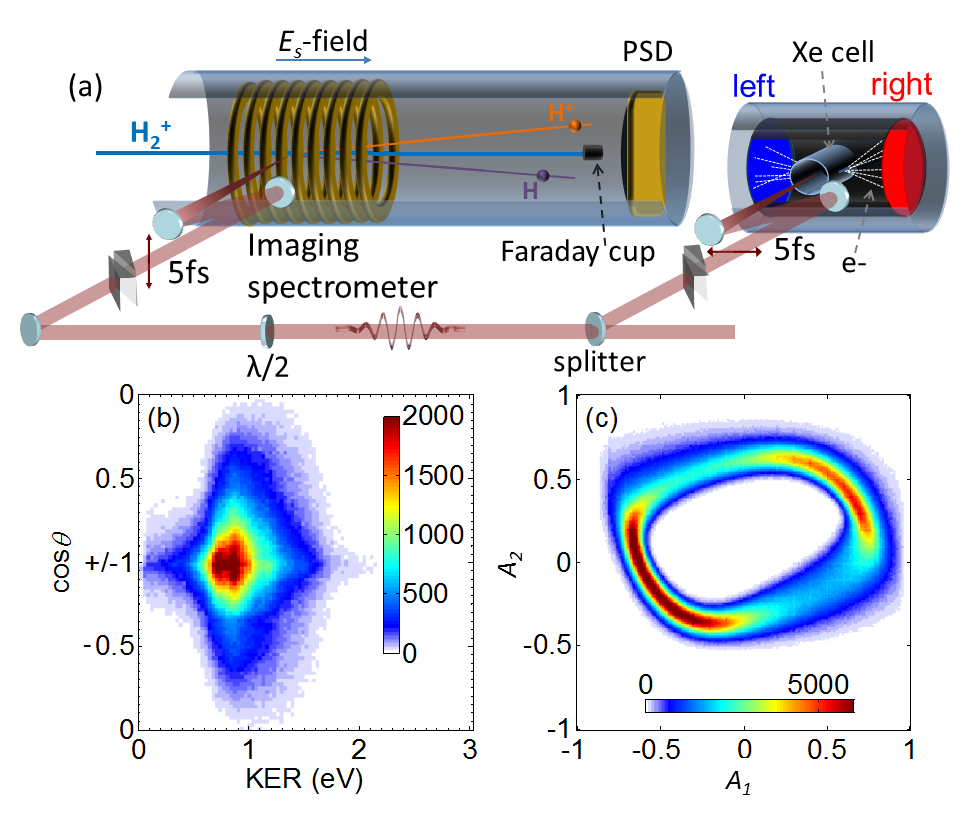}
\caption{(Color online) (a) Schematic view of the experimental setup. The laser beam is split into two arms, and in each arm the dispersion is compensated by a pair of silica wedges. The polarization (indicated by arrows in the path) of the strong arm (80\%) is rotated by a broadband $\lambda$/2 waveplate and is focused into the imaging spectrometer, where it intersects an H$_2^+$ beam. The weak arm is focused into a single-shot stereographic ATI CEP meter \cite{Wittmann:2009,Rathje:2012}, which is used to monitor the CEP at the full repetition rate and CEP tag the molecular data. (b) CEP-integrated dissociation yield as a function of KER and cos\textit{$\theta$}. (c) Measured PAP, from which the CEP is determined (linear color scale).}
\label {figure1}
\end{figure}

Our H$_2^+$ beam is produced in an electron cyclotron resonance (ECR) ion source, accelerated to an energy of 7\,keV, and separated from other ions produced in the ECR by a bending magnet. The ion beam intersects a focused laser beam within an imaging spectrometer with an applied static electric field, $E_s$, that separates the ionic and neutral beam fragments in time, as shown in Fig.\,\ref{figure1}(a). The H$^+$ and H fragments are detected in coincidence on a position- and time-sensitive detector (PSD), while the undissociated molecules are collected in a small Faraday cup. The position and time information allows for the reconstruction of the 3D momenta, from which the KER and angular distributions are evaluated. Figure\,\ref{figure1}(b) shows the dissociation yield as a function of KER and cos$\theta$ where $\theta$ is the angle between the H$^+$ dissociation momentum and the polarization of the laser electric field. More details on the experimental method can be found in Refs.\,\cite{BenItzhak:2005, Wang:2006}.

Pulses of 5\,fs duration with a 730\,nm central wavelength are obtained at a 10\,kHz repetition rate from the PULSAR laser at the J. R. Macdonald Laboratory. The pulses are focused to a peak intensity of 2.5$\times$10$^{15}\,$W/cm$^2$ by an \textit{f}=25\,cm spherical mirror (with a Rayleigh length of about 1.2\,mm). The ion beam crosses the laser 2\,mm in front of the focus, where the peak intensity is 4$\times$10$^{14}\,$W/cm$^2$, in order to take advantage of the larger volume and therefore higher count rate and to minimize the impact of the Gouy phase shift \cite{Lindner:2004}. Under these conditions, ionization remained below 0.2\% of the total signal, making our theoretical approach \cite{Anis:2009, Anis:2012} valid.

\begin{figure} [h!]
\includegraphics[width=0.39\textwidth,trim=0 20 0 10,angle=0,clip=true]{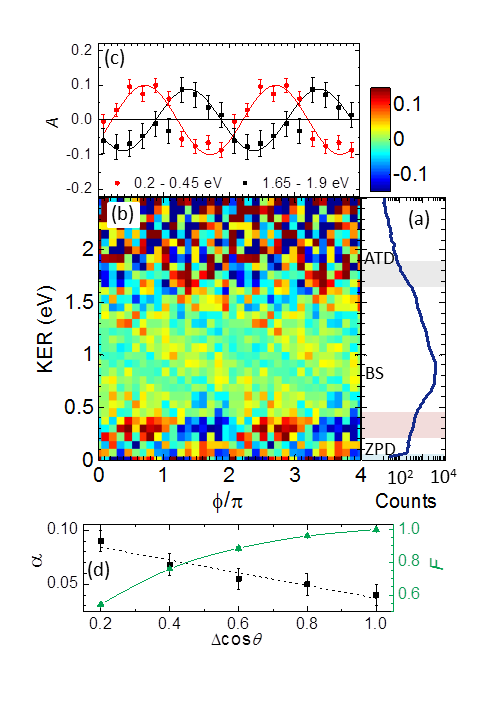}
\caption{(Color online) (a) KER spectrum of H$_2^+$ dissociation by 5\,fs, 4$\times$10$^{14}\,$W/cm$^2$ laser pulses, averaged over $\phi$. The upper (purple) and middle (pink) shaded regions indicate the regions where the highest asymmetries are observed, and the lower (blue) shaded region indicates the losses into the Faraday cup. (b) The corresponding asymmetry map showing the dependence of $A$(KER,$\phi$) on KER and CEP. The data are shown for fragments within a cone, $\Delta$cos$\theta$\,=\,0.2, around the polarization axis. For each KER bin, the asymmetry is shifted to oscillate around zero. (c) The asymmetry parameter integrated over the indicated energy regions, fit to sinusoidal curves (see text). (d) The dependence of the asymmetry amplitude $\alpha$ and fraction of the total counts $F$ within the high energy range on the angular range $\Delta$cos$\theta$ (lines to guide the eye). }
\label{figure2}
\end{figure}

To monitor the CEP of the pulses, a single-shot stereographic above threshold ionization (ATI) phase meter is employed \cite{Wittmann:2009,Rathje:2012}. A broadband beamsplitter picks off 20\% of the laser beam, which is focused by an \textit{f}=25\,cm spherical mirror into the Xe-filled gas cell of the phase meter [see Fig.\,\ref{figure1}(a)]. The electron time-of-flight (TOF) signals are measured by microchannel plate detectors with metal anodes situated to the left and right along the laser polarization. The electron yields in two TOF regions, corresponding to low (region 1) and high (region 2) energies of the measured ATI spectra, are integrated for every laser shot. Then, the respective asymmetries, $A_{1,2}$=$(N_L-N_R)/(N_L+N_R)$, are evaluated where $N_{L(R)}$ is the number of electrons within the TOF gates for the left (right) detector. Plotting $A_1$ and $A_2$ against each other gives rise to a reference parameteric asymmetry plot (PAP), shown in Fig.\,\ref{figure1}(c), which is used to extract the actual CEP up to a constant offset \cite{Sayler:2011}. By simultaneously recording the information from the CEP meter and the molecular dissociation imaging setup, the H$_2^+$ dissociation event is tagged with the CEP of the associated laser pulse. The data presented in Figs.\,\ref{figure1} and \ref{figure2} were taken over 7 hours. In order to reduce the error in the conversion from measured to actual CEP, the data are divided into 20 time-ordered sections and the calculation of the CEP for each section is based on the reference PAP measured during the same time interval \cite{Kuebel:2012}.

Figure\,\ref{figure2}(a) shows the measured KER spectrum (CEP-averaged) for the dissociation of H$_2^+$ into H$^+$\,$+$\,H at a peak laser intensity of 4$\times$10$^{14}\,$W/cm$^2$ within the cone $\theta\,\leq\,36.9^{\circ}$. Several characteristic features of the KER spectrum are labeled in the figure, including the dominant BS \cite{Bucksbaum:1990} region centered around 0.86\,eV with an energy tail that extends to low KER where zero-photon dissociation (ZPD) \cite{Gaire:2011, Posthumus:2000} plays a role, and the above threshold dissociation (ATD) \cite{GiustiSuzor:1990} region at higher energies ($>$1.2\,eV) \cite{Anis:2009}.

For H$_2^+$ dissociation events, the normalized spatial asymmetry is given by
\begin{equation}
A(\mathrm{KER},\phi) = \frac{N_u(\mathrm{KER},\phi)-N_d(\mathrm{KER},\phi)}{N_u(\mathrm{KER},\phi)+N_d(\mathrm{KER},\phi)},
\end{equation}
where $N_{u(d)}($KER$,\phi)$ is the number of H$^+$\,+\,H events with the proton emitted in the up (down) direction, defined by cos$\theta$ being positive (negative). The resulting asymmetry map, $A($KER$,\phi)$, is shown in Fig.\,\ref{figure2}(b). For visualization, the data that is recorded from 0 to 2$\pi$ is duplicated from 2$\pi$ to 4$\pi$.

A clear CEP-dependent asymmetry is present in the very low KER region (0.2\,--\,0.45\,eV) that has not been observed in earlier studies on neutral H$_2$. A second strong CEP-dependent asymmetry is observed at higher KER (1.65\,--\,1.9\,eV). The asymmetries within these two regions are shown in Fig.\,\ref{figure2}(c) as a function of CEP. These oscillatory data were fit to the predicted dominant behavior \cite{Hua:2009, Anis:2012, SI} $A(\phi)=\alpha$cos$(\phi+{\phi}_0)$ --- where $\alpha$ is the asymmetry amplitude and ${\phi}_0$ an offset. For the higher KER region, $\alpha$ is plotted for several cones about the polarization axis, indicated by $\Delta$cos$\theta$ in Fig.\,\ref{figure2}(d), along with $F$, the fraction of the total counts within this energy range. As $\Delta$cos$\theta$ is decreased, the asymmetry amplitude increases. Thus, the cut $\Delta$cos$\theta$=0.2 was chosen for the comparison between experiment and theory. With this choice of angular integration, some weak oscillations between 0.5 and 1.5\,eV having $\alpha$$\approx$0.02 with a KER-dependent offset ${\phi}_0$ (\textit{i.e.} tilt) appear in Fig.\,\ref{figure2}(b).

The origin of the CEP oscillations in the asymmetry can be understood within the theoretical framework proposed by Esry and coworkers \cite{Roudnev:2007, Anis:2009, Hua:2009, Anis:2012, SI}. In this theory, the spatial symmetry is broken through the interference of pathways involving different net numbers of photons that lead to opposite parity states. Starting from an incoherent population of vibrational levels in the 1\textit{s$\sigma_g$} state of H$_2^+$ (generated by electron-impact ionization in the ECR), dissociation of H$_2^+$ occurs through laser-induced coupling to the 2\textit{p$\sigma_u$} state. If any single vibrational level dissociates via different pathways by absorbing and/or emitting different net numbers of photons with the same final energy, then the resulting even and odd nuclear parity states interfere, giving rise to a spatial asymmetry \cite{Roudnev:2007, Anis:2009, Hua:2009, Anis:2012, SI}.  The dominant interference is through pathways where the net photon number differs by one, which leads to the predicted cos($\phi+\phi_0$) dependence of the asymmetry \cite{Hua:2009, Anis:2012, SI} and fits our data well, as shown in Fig.\,\ref{figure2}(c).

\begin{figure} [t]
\begin{center}
\includegraphics[width=0.34\textwidth,trim=0 0 0 0,angle=0,clip=true]{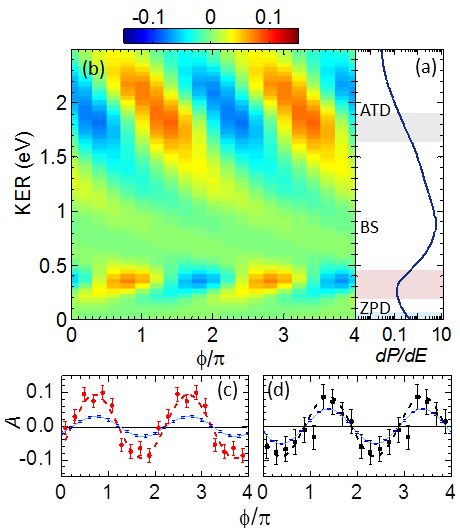}
\end{center}
\caption{(Color online) Calculated (a) $\phi$-averaged dissociation probability, $dP/dE$, as a function of KER with the same shaded regions as in Fig.\,\ref{figure2}(a) and Fig.\,\ref{figure2}(b) $A$(KER,$\phi$) as a function of KER and $\phi$ for the dissociation of H$_2^+$. The asymmetry, averaged over the (c) low and (d) high KER regions re-plotted from Fig. \ref{figure2}(c). The solid light blue lines are theoretical predictions for the same KER regions, with the estimated theoretical error in dark blue \cite{SI2}. The calculations include Franck-Condon factors and intensity averaging.}
\label {figure3}
\end{figure}

The calculated KER spectrum and asymmetry map for 5\,fs Gaussian pulses at 10$^{14}$\,W/cm$^2$ with a central wavelength of 730\,nm are shown in Figs.\,\ref{figure3}(a) and \ref{figure3}(b) \cite{SI2}. Overall, the theory agrees well with the experiment. As $\phi$ is only known up to a constant, arbitrary offset in the experiment, the experimental $\phi$ axes were all shifted by 0.18$\pi$ to match the theory in the high KER region [see Fig.\,\ref{figure3}(d)]. Significantly, after this shift, the experimental and theoretical low-KER asymmetry, shown together in Fig.\,\ref{figure3}(c), are in phase with each other, suggesting that $\phi_0$ is well described by theory. In fact, the asymmetry amplitude is in good agreement for the high KER as well --- it lies within the experimental error bars --- while theory underestimates $\alpha$ by about a factor of three for the low KER. Achieving better quantitative agreement will require further study (both experimental and theoretical) and most likely requires addressing the lower intensity that the theory was limited to \cite{SI2} and any non-Gaussian character of the laser pulse. We know from Ref. \cite{Anis:2012}, for instance, that even a weak prepulse can substantially increase the asymmetry.

\begin{figure} [t]
\includegraphics[scale=0.65,trim=0 0 0 0,angle=0,clip=true]{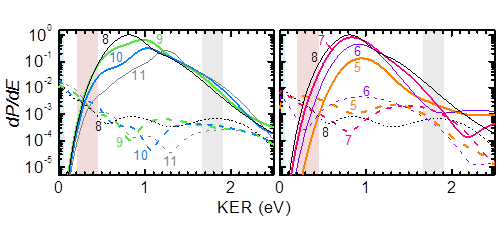}
\caption{(color online) Intensity-averaged dissociation probability, $dP/dE$, as a function of KER for H$_2^+$ in select initial vibrational levels (as indicated), weighted by their Franck-Condon factors. The 2\textit{p$\sigma_u$} (solid lines) and 1\textit{s$\sigma_g$} (dashed lines) dissociation probabilities have comparable magnitudes for the (a) low (shaded pink) and (b) high (shaded purple) KER regions exhibiting high asymmetry.}\label {figure4}
\end{figure}

The agreement between theory and experiment is, however, sufficiently good that we can use the theory to help us identify the important pathways that produce the asymmetry. It must first be clearly understood that the physical observable is the asymptotic relative momentum between an H$^+$ and an H. This outgoing wave atomic scattering state is constructed from a linear combination of the 1\textit{s$\sigma_g$} and 2\textit{p$\sigma_u$} nuclear wavefunctions that takes into account the indistinguishability of the nuclei and includes their spin \cite{Anis:2009, Anis:2012}. Nevertheless, although the experiment sees only the combination of the molecular channels, theory allows for their separate inspection to determine where they overlap to produce the maximum interference. Theory can further provide the molecular channel KER spectra for each initial vibrational state as shown in Fig.\,\ref{figure4}.

Figure\,\ref{figure4} suggests that the pathways contributing to the interference in the high and low energy regions are different in origin. At low KER, ZPD, which is a 2-photon Raman process resulting in the net absorption of 0 photons from the field \cite{Posthumus:2000, Gaire:2011}, interferes with 1-photon BS. The former appears in the 1\textit{s$\sigma_g$} KER spectrum; and the latter, in the 2\textit{p$\sigma_u$} spectrum. Where these two probability densities have comparable magnitude, their interference will have the largest contrast. For the $v=$4\,--\,12 states, a subset of which are shown in Fig.\,\ref{figure4}(a), this confluence occurs precisely in the low KER region where high asymmetry is observed.

In the higher KER range (1.65\,--\,1.9\,eV), the asymmetry likely arises from an interference of 1-photon BS and net 2-photon ATD \cite{McKenna:2008}. The vibrational levels $v$=5\,--\,8 meet the requirements for generating an asymmetry in this region, [see pink-shaded section of Fig.\,\ref{figure4}(b)]. Three-photon ATD contributes at higher KER with a tail that extends to lower KER, but is negligible around 1.7\,eV \cite{McKenna:2012}. Therefore, the 3-photon ATD likely does not play a major role in the observed asymmetry.

The fact that different pathways contribute to the asymmetry at low and high KER also gives a plausible explanation for the clear change in $\phi_0$ seen in Fig.\,\ref{figure2}(c) between the two regions. Moreover, the tilt in the asymmetry can be understood from the fact that at a given intensity and CEP, the relative phase between the interfering pathways also depends on KER.

Since the photon number is not a physical observable, however, and since there are no clearly distinguishable photon peaks --- even in the theoretical molecular channel KER spectra for individual initial vibrational states --- these pathway labels are only approximate. What theory allows us to definitively state is that the net number of photons was even for the 1$s\sigma_g$ channel or odd for 2$p\sigma_u$.

The pathway interference picture can also explain the dependence of the magnitude of the asymmetry on the angles included in the analysis as depicted in Fig.\,\ref{figure2}(d). The 2-photon ATD is a weak channel relative to BS, but it is more aligned with the laser polarization \cite{McKenna:2008}. As the angular range $\Delta$cos$\theta$ around the polarization direction is decreased, ATD becomes more comparable to BS, thus revealing a larger asymmetry. In contrast, when the whole angular distribution is considered, the asymmetry is masked by the strong signal from just the BS channel which does not, by itself, contribute to the asymmetry \cite{Anis:2012}.

One further advantage of our method is that it facilitates the measurement of relative total dissociation yields as a function of CEP. Thus, motivated by Hua and Esry's prediction of a weak CEP effect in the energy-integrated total yield for non-rotating H$_2^+$ in 5.9\,fs, 10$^{14}$\,W/cm$^2$ pulses \cite{Hua:2009}, we searched for but found no discernible dependence of the total yield, integrated over all KER, on CEP within our error bars. This finding is consistent with the present calculations, which give a relative modulation depth of 0.065\%. And, despite our ability to make cuts in the angular distribution to select the molecules that broke while nearly aligned with the laser polarization (limited by post-dissociation rotation \cite{AnisJPB:2009}), intensity averaging apparently washes out any effect. In contrast, Xu \textit{et al.} observed modulation depths of up to 5\% in the H$^+$\,+\,H channel starting from an H$_2$ target, with 6\,fs, $6\times10^{14}$\,W/cm$^2$ pulses \cite{Xu:2013}.

In summary, we have demonstrated CEP effects in the dissociation of an H$_2^+$ molecular ion beam by intense, few-cycle laser fields. Using the one-electron, ionic H$_2^+$ target --- instead of the neutral H$_2$ as in previous experiments --- enabled us to make direct, unambiguous, quantitative comparisons with nearly exact theory. While good in many ways, these comparisons showed that obtaining close quantitative agreement will require further work both theoretically and experimentally. We could show, however, that the mechanisms of the CEP control were generally different from those proposed for H$_2$, but could be understood within a relatively simple --- but exact --- physical picture that applies universally. This picture, in which the CEP controls the relative phase between different dissociative pathways, makes concrete the role that CEP plays in strong-field coherent control. Therefore, we can be more confident in applying it to more complicated systems.

The authors acknowledge A.M. Sayler and J. McKenna for their contributions to earlier experimental efforts, and thank C.W. Fehrenbach for assistance with the ion beam. This work was supported by the Chemical Sciences, Geosciences, and Biosciences Division, Office of Basic Energy Sciences, Office of Science, US Department of Energy. M.F.K. and M.K. acknowledge support by the Max Planck Society and the DFG via grants Kl-1439/3 and Kl-1439/5 and B.J. is supported in part by the Department of Energy Office of Science Graduate Fellowship Program (DOE SCGF), made possible in part by the American Recovery and Reinvestment Act of 2009, administered by ORISE-ORAU under contract no. DE-AC05-06OR23100.

\bibliography{H2pCEPbib3}

\end{document}